\documentclass{cup-hpl}
\usepackage{lipsum}

\usepackage{natbib}

\begin{document}

\newtheorem{theorem}{Theorem}

\shorttitle{Predictive Hydrodynamic Simulations for Laser Direct-drive Implosion Experiments via Artificial Intelligence}
\shorttitle{Predictive Hydrodynamic Simulations for Laser Direct-drive Implosion Experiments via AI}
\shortauthor{Z. Wang, Y. Wang, J. Ma, F. Wu, J. Yan, X. Yuan, Z. Zhang, J. Zhang}

\title{Predictive Hydrodynamic Simulations for Laser Direct-drive Implosion Experiments via Artificial Intelligence}

\author[1,2]{Zixu Wang\textsuperscript{*,}}
\author[1,2]{Yuhan Wang\textsuperscript{*,}}
\author[1]{Junfei Ma\textsuperscript{*,}}
\author[1,3]{Fuyuan Wu}
\author[4]{Junchi Yan}
\author[1,3]{Xiaohui Yuan}
\author[5]{Zhe Zhang}
\author[1,2,3,5]{Jie Zhang\corresp{jzhang1@sjtu.edu.cn and fuyuan.wu@sjtu.edu.cn}}


\address[1]{School of Physics and Astronomy, Shanghai Jiao Tong University, Shanghai 200240, China}
\address[2]{Zhiyuan College, Shanghai Jiao Tong University, Shanghai 200240, China}
\address[3]{State Key Laboratory of Dark Matter Physics, Key Laboratory for Laser Plasmas (MOE) and Collaborative Innovation Center of IFSA, Shanghai Jiao Tong University, Shanghai 200240, China}
\address[4]{School of Artificial Intelligence, Shanghai Jiao Tong University, Shanghai 200240, China}
\address[5]{Beijing National Laboratory for Condensed Matter Physics, Institute of Physics, Chinese Academy of Sciences, Beijing 100190, China}

\begin{abstract}
This work presents predictive hydrodynamic simulations empowered by artificial intelligence (AI) for laser driven implosion experiments, taking the double-cone ignition (DCI) scheme as an example. A Transformer-based deep learning model MULTI-Net is established to predict implosion features according to laser waveforms and target radius. A Physics-Informed Decoder (PID) is proposed for high-dimensional sampling, significantly reducing the prediction errors compared to Latin hypercube sampling. Applied to DCI experiments conducted on the SG-II Upgrade facility, the MULTI-Net model is able to predict the implosion dynamics measured by the x-ray streak camera. It is found that an effective laser absorption factor about 65\% is suitable for the one-dimensional simulations of the DCI-R10 experiments. For shot 33, the mean implosion velocity and collided plasma density reached 195 km/s and 117 g/cc, respectively. This study demonstrates a data-driven AI framework that enhances the prediction ability of simulations for complicated laser fusion experiments.
\end{abstract}


\maketitle

\vspace*{-1.5em}
\renewcommand{\thefootnote}{\fnsymbol{footnote}}
\footnotetext[1]{Zixu Wang, Yuhan Wang, and Junfei Ma contribute equally to this work.}
\renewcommand{\thefootnote}{\arabic{footnote}}

\section{Introduction} \label{sec:Introduction}
There has been many progress toward the realization of laser driven fusion energy in the past decades~\cite{nuckolls1972laser,abu2024achievement,williams2024demonstration,nagatomo2024formation,tikhonchuk2024physics,ditmire2023focused,zhang2020double}. In 2022, the National Ignition Facility (NIF) realized indirect-drive fusion ignition with an energy gain larger than one for the first time~\cite{abu2024achievement}. In 2024, The OMEGA facility achieved the direct-drive hot-spot ignition~\cite{williams2024demonstration}. Recently, the study of the double-cone ignition (DCI) scheme showed promising prospects for the laser fusion energy~\cite{zhang2020double,wu2022machine,lei2024comparison,dai2024diagnosing,liu2024observation}. However, the development of laser fusion still faces a major challenge: the traditional hydrodynamic simulations have poor prediction ability for complicated experiments.

In recent years, artificial intelligence has provided new ideas for fusion research ~\cite{gaffney2024data,humbird2019transfer,humbird2022transfer,gopalaswamy2019tripled,marinak2024numerical,li2023hybrid}. By constructing a surrogate model to replace traditional simulations, AI model can significantly improve computational efficiency~\cite{gaffney2024data}. Moreover, advanced artificial intelligence techniques such as transfer learning and Bayesian inference provide a possibility to bridge the gap between simulations and experiments~\cite{humbird2019transfer,humbird2022transfer,gopalaswamy2019tripled}. However, the existing studies still have some limitations. Firstly, the traditional multilayer perceptron (MLP) architecture has a low efficiency for long sequences of laser waveforms. Secondly, the Latin Hypercube Sampling (LHS) method cannot satisfy the sampling quality requirements in the high-dimensional space.

In this work, we propose a novel artificial intelligence method to improve the prediction ability of hydrodynamic simulations for laser fusion experiments. The method has three characteristics. (1) A deep learning model MULTI-Net is established based on the Transformer architecture to match the time sequence of laser waveform, (2) A high-quality dataset is constructed to increase the prediction ability of the surrogate model by using a physics-informed decoder. (3) The deep learning model is employed to predict laser driven implosion experiments on the SG-II facility.

This paper is organized as follows. Sec.~\ref{sec:Workflow} describes the workflow of the AI-empowered prediction method. Sec.~\ref{sec:MULTI-Net} and Sec.~\ref{sec:Sampling} present the training and improvement of the MULTI-Net model with the physics-informed decoder sampling method. Sec.~\ref{sec:Implosion} applies the MULTI-Net model to predict DCI implosion experiments. Finally, a summary and discussions are given in Sec.~\ref{sec:Conclusion}.

\section{Workflow of AI-Empowered Simulations for Laser Fusion Experiments} \label{sec:Workflow}

\begin{figure*}[htbp!]
\centering
\includegraphics[width=0.99\textwidth]{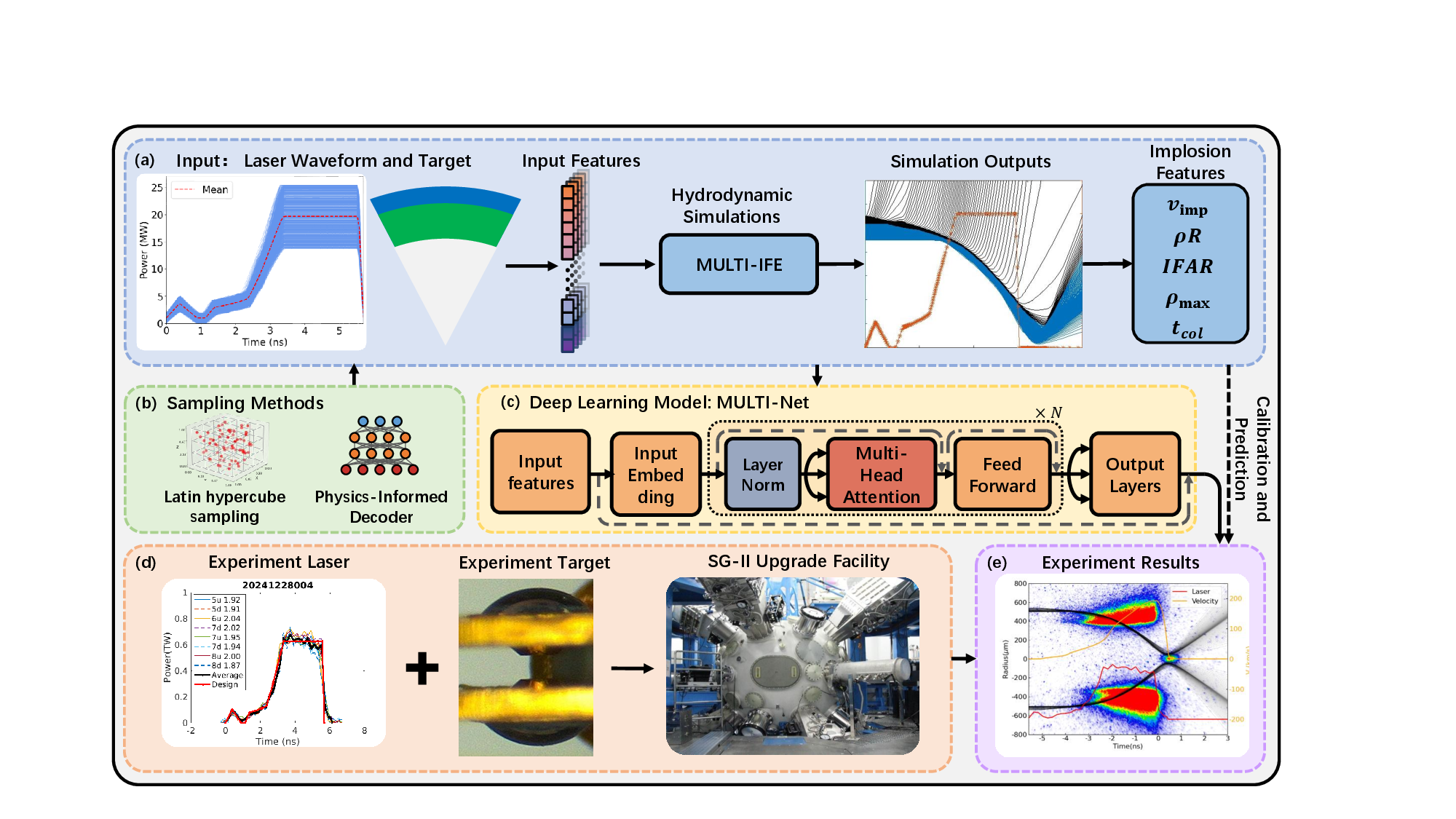}
\caption{\label{fig:1}Workflow of AI-empowered simulations for laser fusion experiments.}
\end{figure*}

Figure~\ref{fig:1} presents a comprehensive workflow for simulating, calibrating, and predicting the DCI experiments in this work. The process begins with hydrodynamic simulations by using the MULTI-IFE program~\cite{ramis2016multi}, where laser waveforms and target parameters serve as the initial conditions (Fig.~\ref{fig:1}(a)). The MULTI-IFE program calculates the corresponding implosion features, including the areal density ($\rho R$), implosion velocity ($V_{\mathrm{mean}}$), in-flight aspect ratio ($IFAR$), maximum density ($\rho_{\mathrm{max}}$), and collision time ($t_{\mathrm{col}}$).

To address the challenges posed by high-dimensional sampling, the study employs advanced sampling methods to generate a training dataset (Fig.~\ref{fig:1}(b)). The deep learning model MULTI-Net based on Transformer architecture is trained with the simulation dataset (Fig.~\ref{fig:1}(c)). The model effectively learns the complex relationships between input and implosion features, significantly reducing computational costs while maintaining high accuracy. The DCI experimental campaign is conducted on the SG-II Upgrade facility (Fig.~\ref{fig:1}(d)). The calibration process enhances the predictive capability of the model by adjusting the effective laser energy absorption (Fig.~\ref{fig:1}(e)).

\section{Deep Learning Model MULTI-Net Based on Transformer} \label{sec:MULTI-Net}
\subsection{The Transformer Architecture of MULTI-Net Model} \label{sec:Architecture}

\begin{figure*}[htbp!]
\centering
\includegraphics[width=0.99\textwidth]{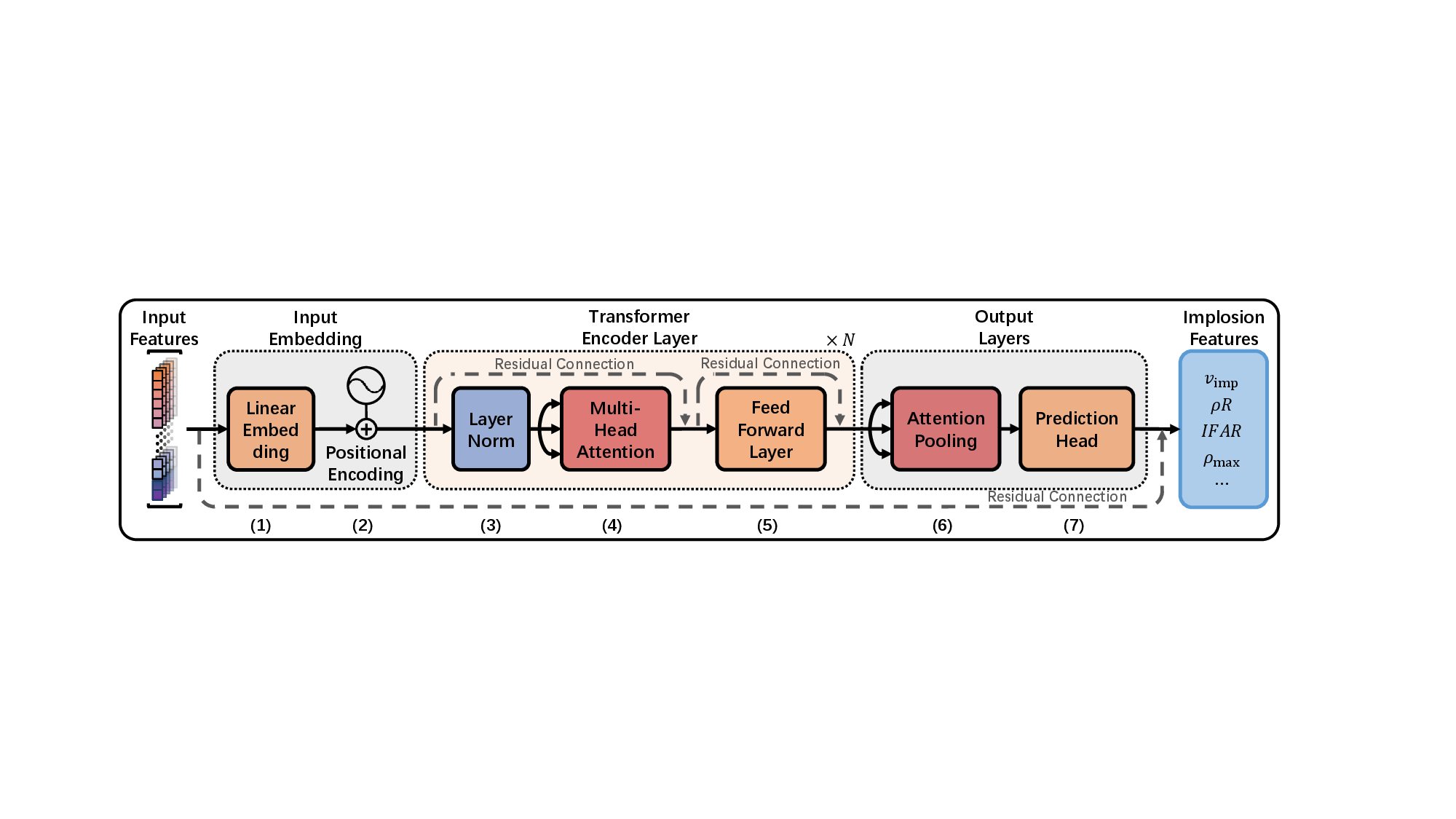}
\caption{\label{fig:2}Sketch of the surrogate model MULTI-Net based on Transformer architecture.}
\end{figure*}

In designing the MULTI-Net model, we first considered the multilayer perceptron (MLP, i.e., fully connected neural network) architecture as the encoder between the input (laser and target parameters) and output features (e.g. implosion velocity and areal density). However, the input parameters such as experimental laser waveforms are generally sequences with hundreds of points, and the MLP model is not suitable for handling high-dimensional input information. Therefore, we replace the MLP module with a module based on the Transformer~\cite{olson2024transformer,vaswani2017attention}, with an encoder-only architecture similar to the large language models such as BERT~\cite{devlin2019bert}.

Compared to the MLP architecture, the multi-head self-attention mechanism in the Transformer architecture can better capture the complex long-range dependencies in the long sequences. Specifically, the MULTI-Net with Transformer architecture consists of the following parts, as shown in Fig.~\ref{fig:2}. (1) Input Embedding: mapping sequence data into high-dimensional representations suitable for Transformer processing. (2) Position Encoding: adding sinusoidal position embeddings that allow the model to understand the sequential nature of series data. (3) LayerNorm: placed before each attention and feed-forward layer, improving gradient flow and training stability compared to post-norm architectures. (4) Multi-Head Self-Attention: capturing complex dependencies and patterns within the series data by computing attention across multiple representation subspaces in parallel. (5) Feed Forward Layer: processing the latent features with fully connected layers, also with pre-norm and residual connection applied. Repeat the Transformer encoder layer consisting of (3)-(5) $N$ times, with residual connections to improve the stability of deep model training. (6) Attention Pooling: aggregating the variable-length sequence into a fixed-size global representation with self-attention mechanism. (7) Prediction Head: transforming the global representation into the final regression outputs with a two-layer MLP, with input residual added.

\begin{figure}[bp!]
\centering
\begin{minipage}{0.49\textwidth}
\centering
\includegraphics[width=0.98\textwidth]{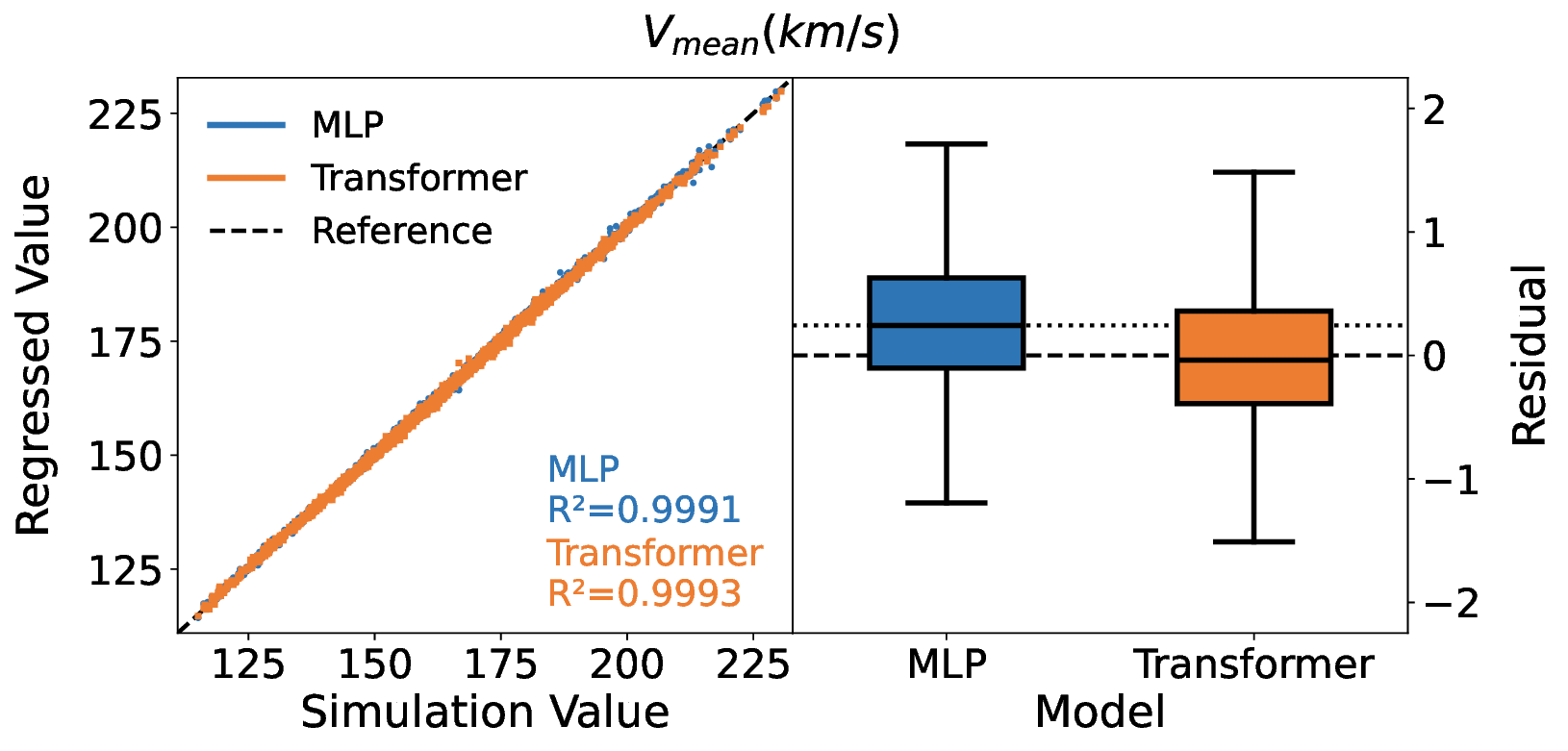}
\end{minipage}
\qquad

\begin{minipage}{0.49\textwidth}
\centering
\includegraphics[width=0.98\textwidth]{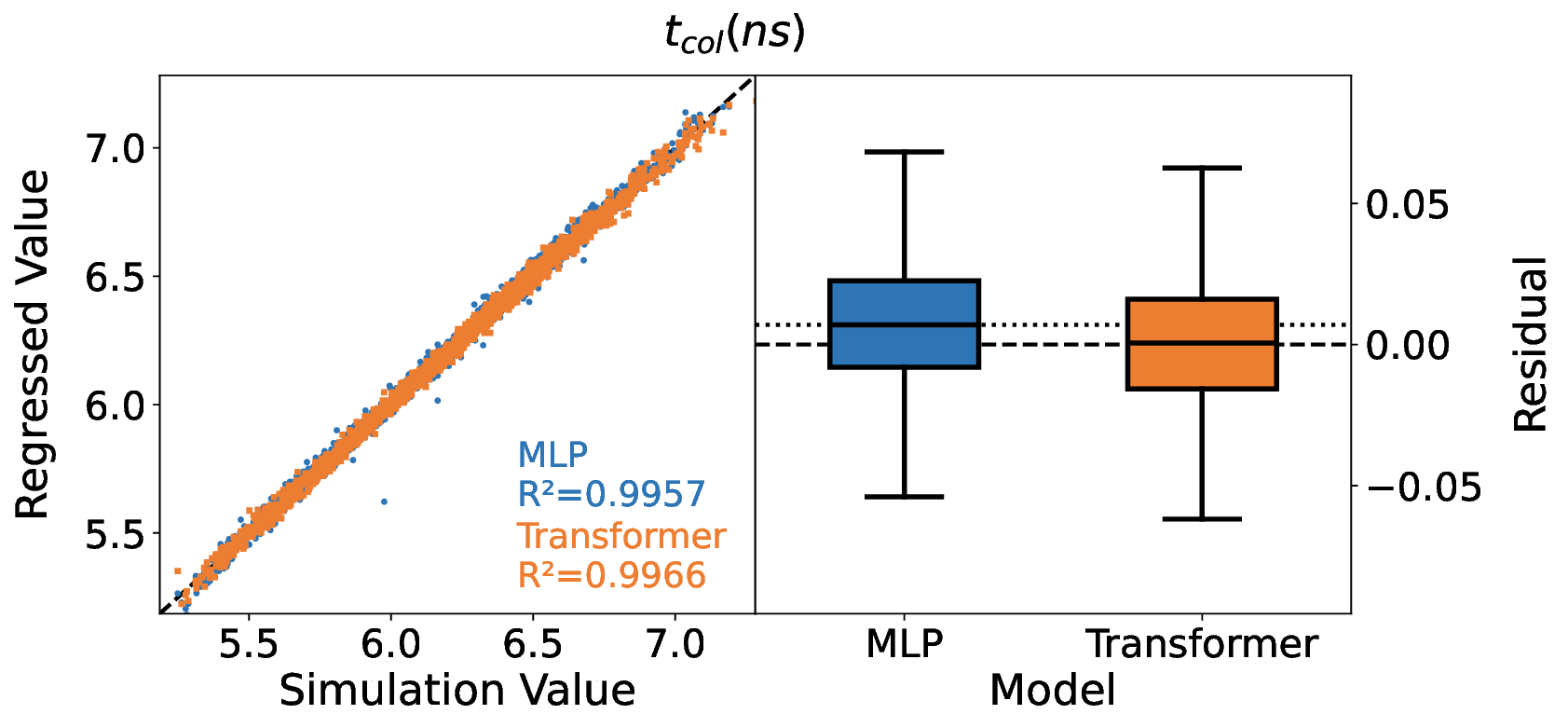}
\end{minipage}
\caption{\label{fig:3}Comparison of the regress ability of the MULTI-Net model with MLP and Transformer architecture for implosion velocity and collision time. The simulated and regressed values are plot with reference line $y=x$ . The boxplots display the interquartile range with the median line.}
\end{figure}

\subsection{The Regress Performance of MULTI-Net Model} \label{sec:Regress}
MULTI-Net's performance is affected by the quality of the training dataset, which should reflect a reasonable probability distribution of the data. While grid sampling is theoretically ideal, it becomes infeasible in high-dimensional spaces due to exponential sample growth. To address this, we employ Latin Hypercube Sampling (LHS) to balance the efficiency and uniformity~\cite{mckay2000comparison}. The total size of the dataset includes 20000 samples, which is divided into 80\% for training, 10\% for testing, and 10\% for validation. The target has a fixed outer radius of 550 \textmu m. The target thickness ranges from 50 \textmu m to 110 \textmu m. The laser energy varies from 16 kJ to 32 kJ.

The structure of the MLP is 101-128-256-128-64-9, whose input includes 100 laser power points and 1 target layer thickness. The architecture of the Transformer has the model dimension $d_{\mathrm{model}}=64$, number of attention heads $n_{\mathrm{head}}=8$, forward dimension $d_{\mathrm{forward}}=128$, and number of Transformer Encoder Layers $N=4$. The total number of both architectures is about 160k.

In order to evaluate the performance of different architectures, we chose two of all implosion features to visualize, namely: the mean velocity ($V_{\mathrm{mean}}$) and collision time ($t_{\mathrm{col}}$). The model predicts the implosion features by learning the time-related laser power sequence. Figure~\ref{fig:3} shows the prediction results of MULTI-Net based on MLP and Transformer architectures with the same number of model parameters. It is shown that the surrogate model based on Transformer architecture exhibits better prediction performance ($R^2$ and residual) for both implosion features. In particular, the regression results of Transformer have a better distribution compared to MLP, with the median of residual reduced by 88.2\% on average, and Transformer achieves this with 10k fewer parameters.

\section{Improving Dataset Quality with Physics-Informed Sampling} \label{sec:Sampling}
\subsection{The Physics-Informed Decoder Sampling Method} \label{sec:PID}
\begin{figure*}[htbp!]
\centering
\includegraphics[width=0.75\textwidth]{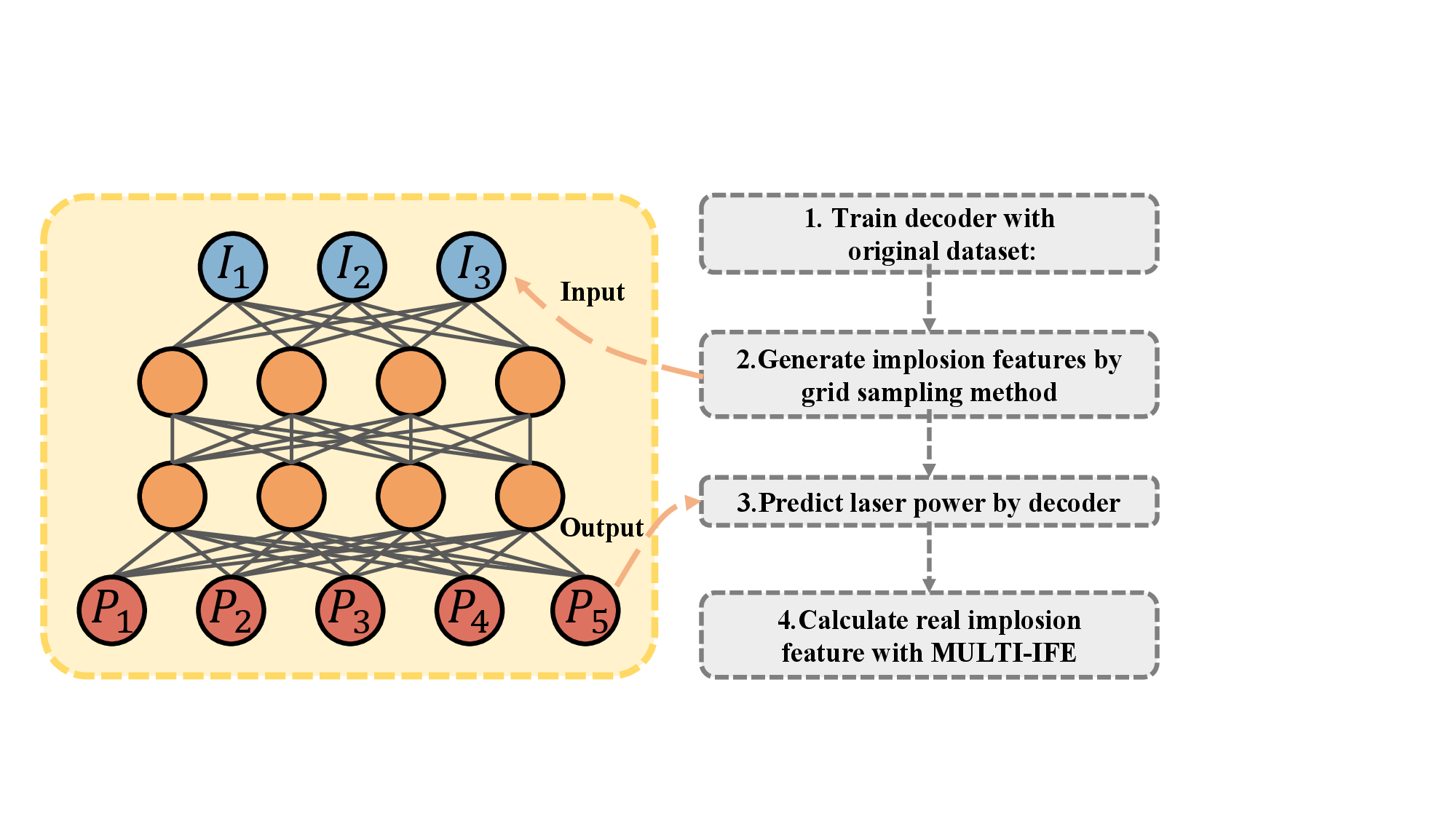}
\caption{\label{fig:4}Sampling method based on Physical-Informed Decoder (PID).}
\end{figure*}

In laser fusion experiments, the experimental data could contain hundreds of freedoms. Through preliminary analysis, we found that interpolating the laser waveform to 100 dimensions (which implicitly contains an equal time interval of $\Delta t = 0.575$ ns in this work) can balance the computational accuracy and cost. Normally, it is extremely difficult to sample efficiently in 100 dimensions. Latin hypercube sampling may inefficiently allocate computational resources to unimportant dimensions. This section realizes a novel sampling method with the physics-informed decoder (PID) by explicitly sampling in the implosion feature space, as shown in Fig.~\ref{fig:4}.

The proposed PID method employs a decoder with a layer configuration of 9-64-64-100. The decoder is first trained to reconstruct laser waveforms from implosion features by using the original dataset. Secondly, implosion features are sampled by the traditional LHS method. Then, the decoder generates laser power waveforms with the input of sampled implosion features. Finally, these generated laser waveforms are employed to compute real implosion features with the code MULTI-IFE.

While traditional sampling methods typically exhibit non-uniform distributions in feature space, the PID sampling method can efficiently generate near-uniform distributions in feature space. This approach allocates more sampling resources to dimensions of particular interest, making it more effective for training on small datasets. It should be noted that the inherent inverse mapping from low-dimensional to high-dimensional parameters of the PID method creates stability constraints. To ensure robustness, the sampling ranges of the PID dataset should be strictly limited within the range of the original dataset.

\subsection{Improvement of MULTI-Net Performance with PID Sampling} \label{sec:Improvement}

In order to evaluate the advantage of the Physics-Informed decoder, this study has made some comparisons. The Latin Hypercube sampling (LHS) improves the sampling efficiency while maintaining the spatial coverage, but it still fails to solve the problem of the high-dimensional exponential explosion. The Physics-Informed Decoder (PID) well solves the disadvantage of the lack of physical information in LHS. While the sizes of the dataset are controlled to 20000, the PID dataset combined 10000 samples from the LHS dataset (original dataset mentioned above) and 10000 samples from PID.

\begin{figure}[bp!]
\includegraphics[width=0.48\textwidth]{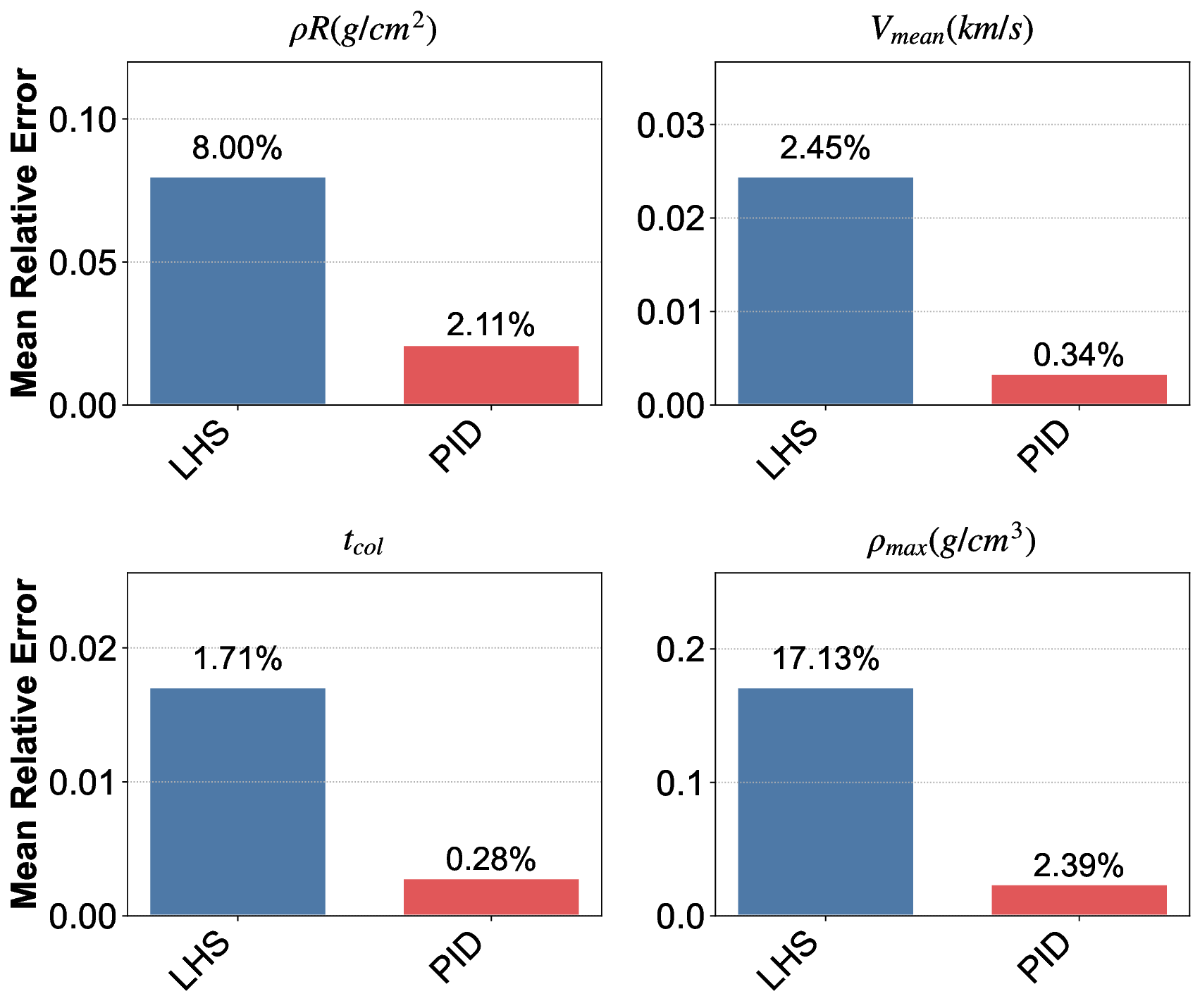}
\caption{\label{fig:5}Prediction ability of the MULTI-Net model with the validation dataset sampled by different sampling methods.}
\end{figure}

Fig.~\ref{fig:5} shows the prediction ability of the MULTI-Net surrogate model on the validation set. The MULTI-Net model is trained on datasets of the same size with the same number of model parameters. It is shown that the models using the PID datasets show better generalization ability, especially for areal density $\rho R$, mean velocity $V_{\mathrm{mean}}$, collision time $t_{\mathrm{col}}$ and max density $\rho_{\mathrm{max}}$, with the PID reducing error by 82.4\% on average. This suggests that the MULTI-Net model is able to make more accurate predictions when trained with the PID dataset.

\section{Prediction of Implosion Dynamics with MULTI-Net} \label{sec:Implosion}
\subsection{Calibration of the Surrogate Model} \label{sec:Calibration}
Many differences in the input of simulations and experiments, such as geometry configuration, boundary conditions, and laser-plasma instability (LPI), cause discrepancies between simulation and experimental results. In similar with the calibrations on NIF~\cite{gaffney2024data}, we modify the laser energy absorption rate to bridge the gap between simulation and experimental results.

This work compares the one-dimensional simulation implosion features $F_{\mathrm{sim}}$ with the experimental implosion features $F_{\mathrm{exp}}$ based on the experimental laser waveforms $L^{\mathrm{exp}}$. The loss function of the calibration for $N$ shots of the experiments is as follows:

\begin{equation*}
\mathcal{L}oss\left(L^{\mathrm{exp}}, L^{\mathrm{eff}}\right) = 
\sum_{N} \sum_{F} \alpha_{F} 
\left\| 
F_{\mathrm{exp}} \left( L^{\mathrm{exp}} \right) 
- 
F_{\mathrm{sim}} \left( L^{\mathrm{eff}} \right) 
\right\|^{2}
\end{equation*}
where $\alpha_{F}$ is the weight coefficient of each implosion feature. The loss function $\mathcal{L}_{\text{loss}}(L^{\text{exp}}, L^{\text{eff}})$ is minimized to find the effective laser waveform $L^{\text{eff}}$ absorbed by the plasma in the simulations. Therefore, the simulation results using the effective laser waveform $L^{\text{eff}}$ would be highly consistent with the experimental results. 

The implosion feature we employed here is the collision time $t_{\text{col}}$, i.e., the moment when the plasma is compressed to the smallest radius. Accurate prediction of the collision moment is important for the DCI experiments because many processes rely on accurate collision moments to obtain optimal results. However, experiments have found that the original MULTI-IFE simulations cannot predict the collision moments accurately enough, so further calibration is needed. In this work, we consider a simple but practical calibration:
\[
L^{\text{eff}} = \eta_{\theta}(L^{\text{exp}})L^{\text{exp}}
\]
where $\theta = (\theta_{1},\theta_{2},\ldots,\theta_{m})$ are parameters to be determined.

\begin{figure}[htbp!]
\includegraphics[width=0.48\textwidth]{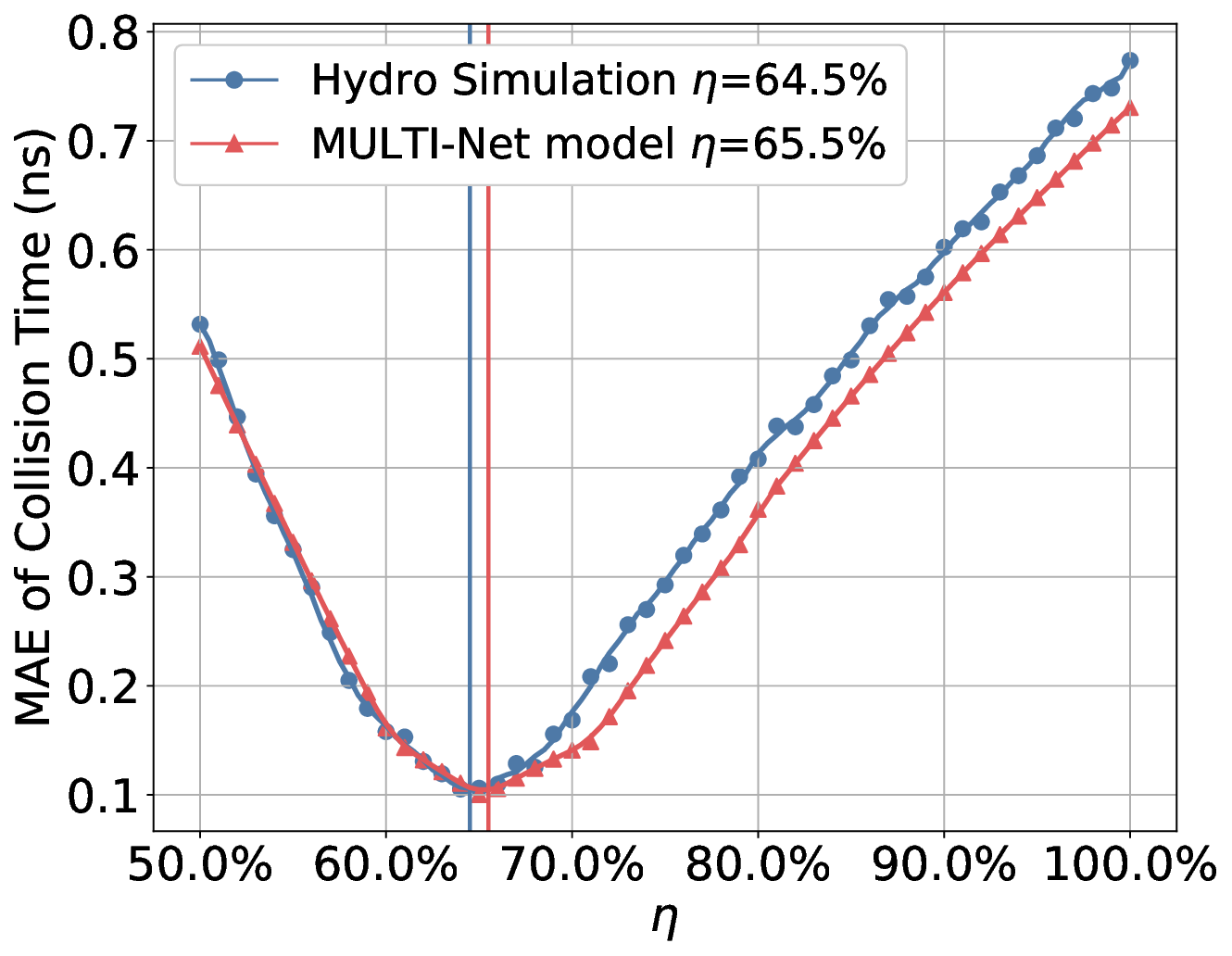}
\caption{\label{fig:6}The effective laser absorption rate calibrated by DCI-R10 experimental results. (MAE: Mean Absolute Error)}
\end{figure}

In the calibration regarding the collision moments, we use 3 shots from the DCI-R10 experiment to perform the calibration, including shots 22, 24, and 37. In the calibration, the energy absorption rate $\eta_{\theta}$ is assumed to be a constant independent of the laser power. We use the \textsc{PID} sampling of the pre-guessed waveforms to train the surrogate model and compare it with original hydrodynamics simulations. It is suggested that the surrogate model greatly accelerates the speed of inference calibration while guaranteeing accuracy. As shown in Fig.~\ref{fig:6}, calibration of three shots suggests that the effective laser absorption rate $\eta_{\theta} = 65\%$ for the MULTI-IFE hydrodynamic simulation and deep learning model \textsc{MULTI-Net}. The error of the collision time $t_{\mathrm{col}}$ in the one-dimensional simulations before calibration is about $0.7\,\mathrm{ns}$, which reduces to $0.1\,\mathrm{ns}$ after calibration. This effective laser absorption rate of $65\%$ indicates that about $35\%$ of the laser energy has to be removed in the one-dimensional implosion. The loss of the laser energy may be caused by the DCI geometry configuration and laser-plasma instability (\textsc{LPI}) in the experiments.

\subsection{Prediction of DCI Experimental Implosion Dynamics} \label{sec:Prediction}
\begin{figure*}[htbp!]
\centering
\includegraphics[width=0.85\textwidth]{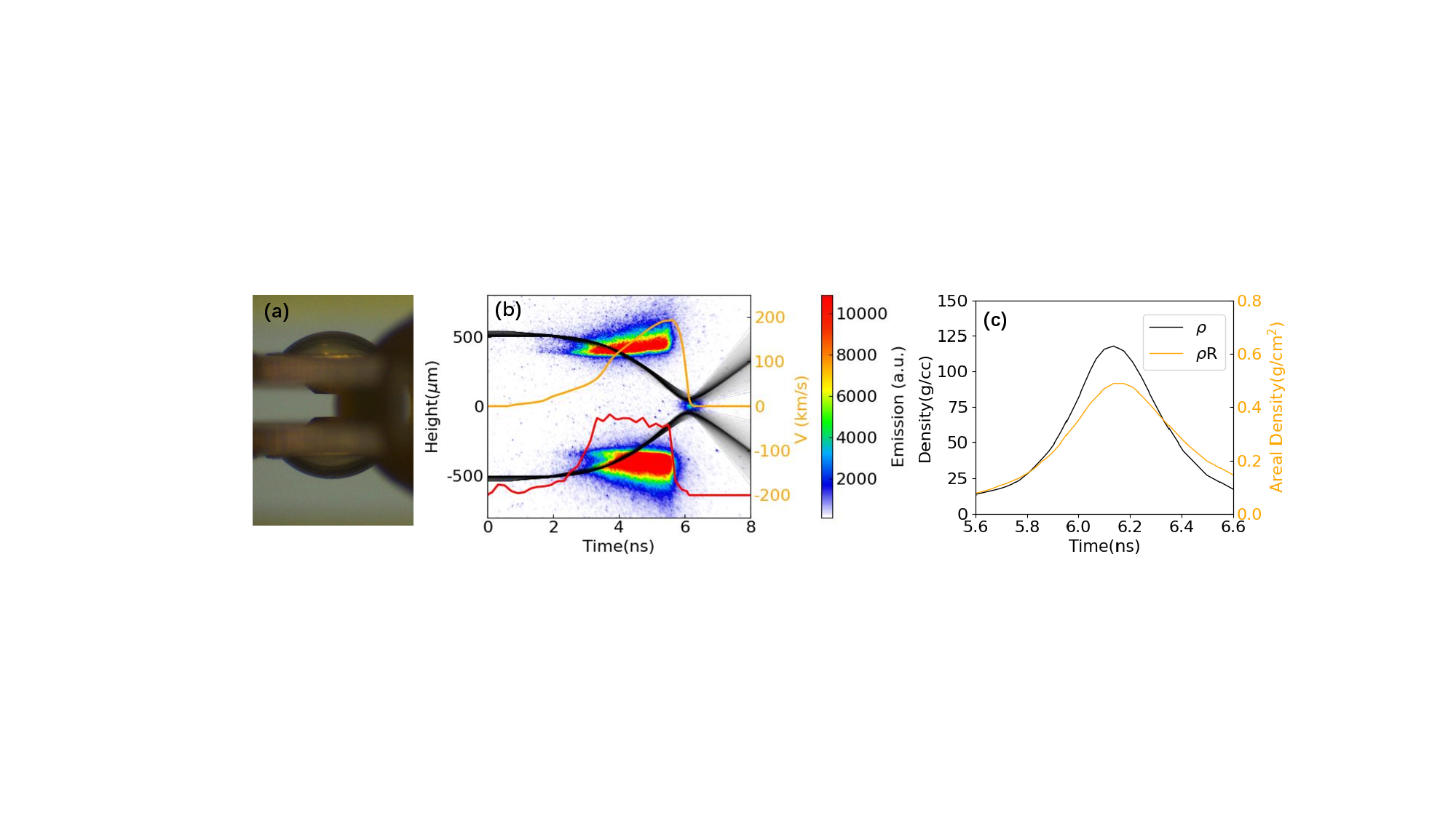}
\caption{\label{fig:7}Prediction of typical DCI implosion dynamics (R10-Shot 33). (a) The photo of double-cone target. (b) The implosion dynamics in experiments and simulations, with plasma implosion trajectories (black lines), implosion velocity (yellow line) and laser power waveform (red line) overlapped on the x-ray emission. (c) The inferred density and areal density during the collision.}
\end{figure*}

For efficient prediction of experimental results, the optimal MULTI-Net model is combined with the calibrated energy absorption rate. Fig.~\ref{fig:7} shows the prediction of a typical shot of the DCI-R10 implosion experiment. The experiment is conducted on the SG-II Upgrade facility located at Shanghai Institute of Optics and Fine Mechanics. The target (Fig.~\ref{fig:7} a) is made of gold cones embedded with plastic shells (C$_8$D$_8$). For shot 33, the CD shell has an outer radius of 560 \textmu m, a thickness of 60 \textmu m and a density of 1.1 g/cc. The CD shells are irradiated by 16 laser beams with a total energy of 28 kJ.

Fig.~\ref{fig:7}(b) presents the x-ray emissions measured by an x-ray streak camera, overlapped with the laser waveform (red line), the implosion trajectories (black lines), and the velocity of the CD shell (yellow line). The x-ray streak camera employs a pinhole as an imaging component, mainly responding to x-rays with energy below 3 keV. The x-ray streak camera has a spatial resolution of $\sim$27 \textmu m and a temporal resolution of 130 ps. It can be observed that the implosion dynamics are well predicted by the simulations after calibration. For example, the experimental collision time (t=6.17 ns) and confinement duration ($\sim$0.4 ns) are captured by the simulation of MULTI-IFE. The simulation trajectories agree well with the acceleration behavior of the CD plasmas reflected by the curve in the X-ray emission (t=2-4 ns). From the viewpoint of power balance, the temperature of ablated corona plasma is dominated by laser heating, thermal conduction, plasma expansion, and radiation cooling. When the laser is shut down, the heating source is lost so that the corona plasmas cool down very fast at the time about 5.69 ns. 

Since the implosion velocity and plasma density are dominated by the plasma motion, we can infer these physical quantities from the simulations. As shown in Fig.~\ref{fig:7}, the implosion velocity of the shell is about 195 km/s, the peak collision density is about 117 g/cc, and the peak areal density is about 0.48 g/cm$^2$. 

Using an absorption rate of 65.5\%, the MULTI-Net model predicts collision time at 6.11 ns with a mean velocity of 190 km/s and an areal density of 0.47 g/cm$^2$. Compared to the uncalibrated results (collision time: 5.55 ns, implosion velocity: 215 km/s, peak areal density: 0.51 g/cm$^2$), the calibrated predictions show significant improvement and align closely with experimental observations. This agreement enhances confidence for the future simulation-experiment synergy.

However, it should be noted that the collision of the plasma jets in the DCI scheme is intrinsically two-dimensional. The actual density and temperature distributions in experiments can only be predicted when two-dimensional simulations are available. In the future, we will try to make more comprehensive predictions with higher dimensional simulations.

\section{Conclusions and Discussions} \label{sec:Conclusion}
This study presents AI-empowered predictive simulations for laser driven implosion experiments. The one-dimensional MULTI-IFE hydrodynamic program is employed to generate datasets for the training of deep learning models. The surrogate model MULTI-Net is constructed based on the Transformer architecture to match the time sequence laser waveforms. In addition, we propose a Physics-Informed Decoder (PID) sampling method to improve the dataset quality compared to Latin hypercube sampling. During the experimental calibration, the transformer-based surrogate model aligns well with the simulation and experimental results. The estimated effective laser absorption rate by the CD plasma is about 65\% in the DCI-R10 experiment campaign. The overall implosion dynamics are predicted by the calibrated simulations, including the plasma collision time, plasma confinement duration, and the shell acceleration behavior. It is found that the typical DCI-R10 implosion experiments have a mean implosion velocity about 195 km/s and a collision density about 117 g/cc.

It should be noted that the whole implosion dynamics of DCI experiments can only be predicted accurately when higher dimensional simulations are available. The MULTI-Net is a part of the multi-modal artificial intelligence model (ZhuRong-I) designed for laser fusion experiments. In the future, we will extend this AI-based framework to broader experimental scenarios, such as the prediction of two-dimensional plasma distribution of the collided plasmas.

\begin{center}\textbf{Acknowledgement}\end{center}

We thank the SG-II Upgrade laser facility operating group and target fabrication team for tremendous help. This work was supported by the Strategic Priority Research Program of Chinese Academy of Sciences (Grant nos. XDA25010100, XDA25051200 and XDA25020200), National Natural Science Foundation of China (No. 12205185), and the IAEA Coordinated Research Project on AI for Accelerating Fusion (No. F13022). The computations in this work were supported by the Center for High Performance Computing at Shanghai Jiao Tong University.


\begin{center}\textbf{References}\end{center}

\end{document}